\documentstyle[12pt]{article}

\begin{document}
\baselineskip 22pt 
\begin{titlepage} 
\begin{center} 
{\Large\bf Rudiments of Dual Feynman Rules for Yang-Mills Monopoles in 
Loop-space} 
\vskip .1cm 
{\large Chan Hong-Mo}\\ 
\vskip .02cm 
{\it Rutherford Appleton Laboratory,\\ 
Chilton, Didcot, Oxon, OX11 0QX, U.K.}\\ 
\vskip .1cm
{\large Jacqueline  Faridani}\\ 
\vskip .02cm 
{\it Department of Physics, University of Toronto,\\ 
60 St.  George St., Toronto, Ontario, M5S 1A7, Canada.}\\ 
\vskip .1cm
{\large Jakov Pfaudler}\\ 
\vskip .02cm 
{\it Department of Theoretical Physics, Oxford University,\\ 
1 Keble Rd., Oxford, OX1 3NP, U.K.}  
\vskip .1cm
{\large Tsou Sheung Tsun}\\ 
\vskip .02cm 
{\it Mathematical Institute, Oxford
University,\\ 
24-29 St.Giles', Oxford, OX1 3LB, U.K.}\\ 
\end{center} 
\begin{abstract}
Dual Feynman rules for Dirac monopoles in Yang-Mills fields are obtained
by the Wu-Yang (1976) criterion in which dynamics result as a consequence 
of the constraint defining the monopole as a topological obstruction in 
the field. The usual path-integral approach is adopted, but using loop-space 
variables of the type introduced by Polyakov (1980).  An antisymmetric 
tensor potential $L_{\mu\nu}[\xi|s]$ appears as the Lagrange multiplier 
for the Wu-Yang constraint which has to be gauge-fixed because of the 
``magnetic'' $\widetilde U$-symmetry of the theory.  Two sets of ghosts are 
thus introduced, which subsequently integrate out and decouple.  The 
generating functional is then calculated to order $g^0$ and expanded in
a series in $\widetilde g$.  It is shown to be expressible in terms of 
a local ``dual potential'' $\widetilde A_\mu(x)$ found earlier, which has
the same propagator and the same interaction vertex with the monopole field
as those of the ordinary Yang-Mills potential $A_\mu$ with a colour charge,
indicating thus a certain degree of dual symmetry in the theory.  For the 
abelian case the Feynman rules obtained here are the same as in QED to all 
orders in $g$, as expected by dual symmetry.
\end{abstract} 
\end{titlepage}

\clearpage

\baselineskip 16pt

\section{Introduction}

It has long been known that monopoles in gauge theories acquire through
their definition as topological obstructions in the gauge field an 
intrinsic interaction with the field.  In fact, in an inspiring paper 
of 1976, Wu and Yang \cite{Wuyang} first showed by a beautiful line of 
argument how the standard (dual) Lorentz equation for a classical point 
magnetic charge could be derived as a consequence of its definition as a 
monopole of the Maxwell field.  Since electromagnetism is dual symmetric, 
it follows that the ordinary Lorentz equation for a point electric charge 
can also be derived by considering the latter as a monopole of the dual 
Maxwell field.  Moreover, it can be seen that this approach for deriving 
the interactions of monopoles, which we shall henceforth refer to as the 
Wu-Yang criterion, is in principle not restricted alone to electromagnetism.  
Indeed, having been supplemented by some technical development necessary 
for its implementation, the method has since been generalized to monopole
charges in nonabelian Yang-Mills theories \cite{lubkin,wuyang75,coleman}, 
not only for classical point particles but also for Dirac particles, giving 
respectively the Wong and the Yang-Mills-Dirac equations or their respective 
generalized duals as the result. \cite{Chanstsou,Chantsou,Chanftsou}

All this work so far on the Wu-Yang criterion, however, has been restricted 
to the classical field level.  The purpose of the present paper is to begin 
exploring the dynamics of nonabelian monopoles at the quantum field level 
as implied by the same Wu-Yang criterion.  We shall start by attempting to 
derive some rudiments of the ``dual Feynman rules'' in this approach.

One purpose of this exercise is to compare the Feynman rules so derived for 
(colour) monopoles with those for (colour source) charges of the standard 
approach.  Although it has recently been shown that nonabelian Yang-Mills 
theory possesses a generalized dual symmetry in which monopoles and sources 
play exact dual roles \cite{Chanftsou1}, so that the dynamics of (colour) 
charges derived using the Wu-Yang criterion when they are considered as 
monopoles of the field is the same as that of the usual Yang-Mills dynamics 
when these charges are considered as sources, this result is again known 
to hold so far only at the field equation level.  On the other hand, the 
exciting fully quantum investigation program on duality initiated by Seiberg 
and Witten and extended by many others \cite{Seiten,Seiberg,Witten,Aharony} 
applies at present strictly only to supersymmetric theories in a framework 
in which the Wu-Yang criterion plays no role, and is for these reasons not 
yet very helpful to the questions raised in the present paper.  The crucial 
point is the existence of the dual potential which is guaranteed only by the 
equation of motion obtained by extremizing the action and thus need no longer 
hold in the quantum theory when the field variables move off-shell.  It is 
therefore interesting to explore whether this generalized dual symmetry 
breaks down at the quantum field level and if so in what way.  Furthermore, 
even if the presently known generalized dual symmetry is eventually seen 
to apply also at the quantum field level, as seems to us possible, we believe 
that our investigation here is still likely to prove useful in future for 
attacking the ultimate problem of both (colour) electric and magnetic charges 
interacting together with the Yang-Mills field.

Another purpose of this work is mainly of technical interest, namely to 
examine how Feynman integrals work in loop space.  As is well-known, the 
loop space approach to gauge theory is attractive in that it gives in 
principle a gauge independent description in terms of physical observables, 
in contrast to the standard description in terms of the gauge potential 
$A_\mu(x)$.  A grave drawback of the loop space approach, however, is the 
high degree of redundancy of loop variables which necessitates the imposition 
on them of an infinite number of constraints to remove this redundancy, making 
thus the whole approach rather unwieldy.  For the problem of nonabelian 
monopoles, on the other hand, it turns out that it pays for various reasons 
to work in loop space, and a set of useful tools has been developed for the 
purpose. \cite{Chanstsou,Chantsou,Chanftsou} In fact, it was only by means of
these loop space tools that the results quoted above on nonabelian monopoles 
at the classical field level have so far been derived.  We are therefore keen 
to investigate how these tools apply to Feynman integrals at the quantum 
field level, the understanding of which, we think, may contribute towards 
the future utilization of the loop space technique as a whole.

That the definition of a charge as a topological obstruction in a field 
should imply already an interaction between the charge and the field is
intuitively clear, because the presence of a charge at a point $x$ in 
space means that the field around that point will have a certain
topological configuration.  When that point moves, therefore, the field 
around it will have to re-adapt itself so as to give the same topological 
configuration around the new point.  Hence, it follows that there must be 
a coupling between the coordinates of the charge and the variables 
describing the field, or in other words in physical language, an 
``interaction'' between the charge and the field.

The Wu-Yang criterion enframes the above intuitive assertion as follows.
One starts with the free action of the field and the particle, which one
may write symbollically as:
\begin{equation}
{\cal A}^0 = {\cal A}^0_F + {\cal A}^0_M,
\label{calA0}
\end{equation}
where ${\cal A}^0_F$ depends on only the field variables and ${\cal A}^0_M$
on only the particle variables. If the variables are regarded as independent,
then the field is completely decoupled from the particle.  However, by
specifying that the particle is a topological obstruction of the field, 
one has imposed a constraint on the system in the form of a condition 
relating the field variables to the particle variables.  Hence, for example, 
if one extremizes the free action (\ref{calA0}) subject to this constraint, 
one obtains not free equations any more but equations with interactions between 
the particle and the field.  Indeed, it was in this way that the Wu-Yang 
criterion has been shown to lead to the Lorentz-Wong and Dirac-Yang-Mills 
equations for respectively the classical and Dirac charge. \cite{Wuyang,
Chanstsou,Chanftsou}

For the quantum theory, the equations of motion will not be enough.  One 
will need instead to calculate Feynman integrals over the field and particle 
variables with the exponential of the action (\ref{calA0}) above as a 
weight factor.  If the variables are regarded as independent and integrated 
freely with respect to one another, then we have again a free decoupled 
system, but since the particle and field variables are here related by the 
constraint specifying that the particle carries a monopole charge, the 
resulting Feynman integrals will involve interactions between the particle 
and the field.  Our aim in this paper then is just to evaluate some such 
Feynman integrals to see what sort of interactions will emerge.

Let us now be specific and consider an $su(2)$ Yang-Mills field with a
Dirac particle carrying a (colour) magnetic charge.  The free action in 
that case is:\footnote{Although given explicitly only for $su(2)$, our 
results are trivially generalizable to all $su(N)$ theories.  In our 
convention for $su(2), B = B^iT_i, T_i = \tau_i/2, Tr B = 2 \times$ 
sum of diagonal elements, so that $Tr (T_iT_j) = \delta_{ij}$.  Our metric is
$g_{\mu\nu} = diag(1,-1,-1,-1)$.}
\begin{equation}
{\cal A}^0_F = -\frac{1}{16 \pi} \int d^4x Tr\{F_{\mu\nu}(x) F^{\mu\nu}(x)\},
\label{calA0F}
\end{equation}
with:
\begin{equation}
F_{\mu\nu}(x) = \partial_\nu A_\mu(x) - \partial_\mu A_\nu(x)
   + ig [A_\mu(x), A_\nu(x)],
\label{FmunuinA}
\end{equation}
for the field in terms of the gauge potential $A_\mu(x)$ as variable, and:
\begin{equation}
{\cal A}^0_M = \int d^4x \bar{\psi}(x) (i \partial_\mu \gamma^\mu - m) \psi(x),
\label{calA0M}
\end{equation}
for the particle in terms of the wave function $\psi(x)$ as variable.

In the presence of monopoles, however, $A_\mu(x)$ has to be patched, which 
makes it rather clumsy to use in this problem.  For this reason, it was 
found convenient in all previous work on the classical theory \cite{Chanstsou,
Chantsou,Chanftsou} to employ as field variable instead the Polyakov variable 
$F_\mu[\xi|s]$ \cite{Polyakov} defined as:
\begin{equation}
F_\mu[\xi|s] = \frac{i}{g} \Phi[\xi]^{-1} \delta_\mu(s) \Phi[\xi],
\label{Fmuxis}
\end{equation}
for:
\begin{equation}
\Phi[\xi] = P_s \exp ig \int_0^{2\pi} ds A_\mu(\xi(s)) \dot{\xi}^\mu(s),
\label{Phixi}
\end{equation}
where $\Phi[\xi]$ is the holonomy element for the loop parametrized by the 
function $\xi$ of $s$ for $s = 0 \rightarrow 2 \pi$ with $\xi(0) = \xi(2\pi) 
= P_0$, or in other words, maps of the circle into space-time beginning 
and ending at the fixed reference point $P_0$, and $\delta_\mu(s)
= \delta/\delta \xi^\mu(s)$ is the functional derivative with respect to
$\xi^\mu$ at $s$.  In terms of $F_\mu[\xi|s]$ as variable, the free action 
of the field now reads as:
\begin{equation}
{\cal A}_F^0 = \int \delta \xi ds a_\xi(s) Tr \{F_\mu[\xi|s] F^\mu[\xi|s]\},
\label{calA0FL}
\end{equation}
where:
\begin{equation}
a_\xi(s) = -\frac{1}{4\pi{\bar N}} \dot{\xi}(s)^{-2},
\label{axis}
\end{equation}
with $\dot{\xi}^\mu(s)$ being the tangent to the loop $\xi$ at $s$ and
${\bar N}$ an (infinite) normalization factor defined as:
\begin{equation}
{\bar N} = \int_0^{2\pi} ds \int \prod_{s' \neq s} d^4\xi(s'),
\label{Nbar}
\end{equation}
and where the integral is to be taken over all parametrized loops\footnote{We 
note that parametrized loops $\xi$ being by definition just 
functions of $s$, integrals over $\xi$ are just ordinary functional 
integrals, which is in fact one reason why we prefer to work with 
parametrized loops rather than the actual loops in space-time.} and over 
all points $s$ on each loop.

The action (\ref{calA0}), with ${\cal A}_F^0$ as given in (\ref{calA0FL}) 
and ${\cal A}_M^0$ as given in (\ref{calA0M}), is subject to constraints
on two counts.  First, the variables $F_\mu[\xi|s]$, as already noted, are 
highly redundant as all loop variables are and have to be constrained so
as to remove this redundancy.  Second, the stipulation that the particle
represented by $\psi(x)$ should correspond to a monopole of the field
implies that $\psi(x)$ must be related to the field variable $F_\mu[\xi|s]$
by a topological condition representing this fact.  The beauty of the loop
space formalism is that both these constraints are contained in the single
statement:
\begin{equation}
G_{\mu\nu}[\xi|s] = -4\pi J_{\mu\nu}[\xi|s],
\label{Gausslaw}
\end{equation}
where:
\begin{equation}
G_{\mu\nu}[\xi|s] = \delta_\nu(s) F_\mu[\xi|s] - \delta_\mu(s) F_\nu[\xi|s]
   + ig [F_\mu[\xi|s], F_\nu[\xi|s]],
\label{Gmunuxis}
\end{equation}
is the loop space curvature with $F_\mu[\xi|s]$ as connection, and 
$J_{\mu\nu}[\xi|s]$ is essentially just the (colour) magnetic current
carried by $\psi(x)$, only expressed in loop space terms, the explicit
form of which will be given later but need not at present bother us.\footnote
{Strictly speaking, to remove completely their redundancy, the variables 
$F_\mu[\xi|s]$ are required to have vanishing components along the direction 
of the loop $\xi$, which ``transversality condition'' has in principle to 
be treated as an additional constraint on the system. \cite{Chanstsou} This 
constraint is however easily handled though giving added complications.  The 
calculations reported in this paper have actually been done taking full 
account of transversality but since the result is the same, the arguments 
are not given here for the sake of a simpler presentation.  For details, 
see ref. \cite{Faridani,Pfaudler}.}

That being the case, the Wu-Yang criterion then says that the dynamics of
the monopole interacting with the field is already contained in the constraint
(\ref{Gausslaw}).  Indeed, it was by extremizing the `free' action (\ref{calA0})
under this constraint (\ref{Gausslaw}) that in our earlier work the (dual) 
Yang-Mills equations of motion for the monopole have been derived.  To extend
now the considerations to the quantum theory, we shall need to evaluate
Feynman integrals over the variables $F_\mu[\xi|s]$ and $\psi(x)$, but subject
again to the constraint (\ref{Gausslaw}).  Thus, the partition function of the
quantum theory would be of the form:
\begin{equation}
Z = \int \delta F \delta \psi \delta \bar{\psi} \exp i{\cal A}^0 
  \prod_{\mu,\nu,[\xi|s]} \delta \{G_{\mu\nu}[\xi|s] + 4\pi J_{\mu\nu}[\xi|s]\}.
\label{Z1}
\end{equation}
Equivalently, writing the $\delta$-functions representing the constraint as
Fourier integrals, we have:
\begin{equation}
Z = \int \delta F \delta \psi \delta \bar{\psi} \delta L \exp i {\cal A},
\label{Z2}
\end{equation}
with:
\begin{equation}
{\cal A} = {\cal A}^0 
   + Tr\{L^{\mu\nu}[\xi|s] (G_{\mu\nu}[\xi|s] + 4\pi J_{\mu\nu}[\xi|s])\}.
\label{calA}
\end{equation}

Since basically the only functional integral we can do is the Gaussian, the
standard procedure is to expand into a power series all terms of higher order 
in the exponent of the integrand and perform the integral power by power in 
the expansion.  We shall follow here the same procedure.  However, in contrast 
to the usual cases met with in quantum field theory, there are in the 
exponent of the integrand in (\ref{Z2}) two terms of order higher than the 
quadratic, namely one coming from the commutator term of $G_{\mu\nu}[\xi|s]$ 
in (\ref{Gmunuxis}) which is proportional to the Yang-Mills coupling $g$, and 
the other coming from $J_{\mu\nu}[\xi|s]$ which is proportional to the colour 
magnetic charge $\tilde{g}$.  The result of the expansion would thus be a 
double power series in $g$ and $\tilde{g}$.  In view of the fact that $g$ 
and $\tilde{g}$ are related by the Dirac quantization condition which means 
usually that if one is small then the other will be large, we can normally 
regard such a double series only as a formal and not as a perturbation 
expansion.  Only in certain special circumstances can one see it leading 
possibly to an approximate  perturbative method.  For example, for gauge 
group $SU(N)$, the Dirac condition reads as:
\begin{equation}
g \tilde{g} = 1/2N,
\label{Diracqc}
\end{equation}
with an additional factor $N$ compared with the standard Dirac condition for
electromagnetism.  Thus if the effective gauge symmetry is continually 
enlarged so that $N \rightarrow \infty$ as energy is increased as some believe 
it may, then in principle both $g$ and $\tilde{g}$ can be asymptotically 
small.  A case of perhaps more practical interest is quantum chromodynamics 
with $N = 3$ where for $Q$ ranging from 3 to 100 GeV, phenomenological values 
quoted for $\alpha_s$ run from about .25 to .115 \cite{Partdata}.  This 
corresponds to $g$, say, running from about 1/2 to 1/3 \footnote{Notice 
that the coupling $g$ occurring in the Dirac condition (\ref{Diracqc}) is 
the so-called unrationalized coupling related to $\alpha_s$ by $g^2 = \alpha_s$ 
without a factor $4\pi$.} and implies by (\ref{Diracqc}) that the dual coupling 
$\tilde{g}$ runs also in the same range, namely from 1/3 to 1/2.  Thus, if 
we accept, as is at present generally accepted, that the expansion in $g$ 
gives a reasonable approximation, then it is not excluded that a parallel 
expansion in $\tilde{g}$ can also do so.  However, as far as this paper is 
concerned, we treat the double expansion in $g$ and $\tilde{g}$ merely as 
a formal means of generating Feynman diagrams, the study of which only is 
our immediate purpose.

The expansion having been made, the evaluation of the remaining Gaussian 
integrals then proceeds along more or less conventional lines apart from 
two complications.  First, as was shown in an earlier work \cite{Chanftsou},
the theory possesses now an enlarged gauge symmetry, from the original 
$SU(N)$ doubled to an $SU(N) \times SU(N)$ where the second $SU(N)$ has
a parity opposite to that of the first and is associated with the phase
of the monopole wave function $\psi(x)$.  Under this second $SU(N)$ symmetry
the Lagrange multiplier $L_{\mu\nu}[\xi|s]$ occurring in the integral (\ref{Z2})
transforms as an antisymmetric tensor potential of the Freedman-Townsend type 
\cite{Freedsend} and has thus to be gauge-fixed using the technology given
in the literature for such tensor potentials. \cite{Namazie,Siegel,Fosco}  
Second, the field variables $F_\mu[\xi|s]$ and $L_{\mu\nu}[\xi|s]$ being 
themselves functionals (i.e. functions of the parametrized loops $\xi$ 
which are functions of $s$), extra care has to be used in defining functional
operations, such as the Fourier transform, of the field  quantities.  Apart 
from these complications, the calculations are otherwise fairly straightforward.

In this paper, we have carried the calculation only to order $g^0$. 
Although there is in principle no great difficulty apart from complication 
to carry some of the calculation to higher orders in $g$, and we have done so 
for exploration, the expansion cannot yet be carried out systematically until 
some basic questions are resolved.  Nevertheless, even the simple examples 
we have calculated are sufficient to demonstrate several interesting facts.  
First, that it is possible, though unwieldy, to calculate Feynman diagrams in 
loop space.  Secondly, that the Wu-Yang criterion does yield specific rules for
evaluating Feynman diagrams of monopoles interacting with the field.  Thirdly,
that the result so far is dual symmetric to the standard interaction of a 
colour (electric or source) charge.  We are therefore hopeful that these, 
albeit yet strictly limited, results will give at least a foothold to serve 
as a base for extending the exploration further.

\setcounter{equation}{0}
\section{Preliminaries, Gauge-Fixing and Ghosts}

We begin by quoting from earlier work the form of the monopole (or colour 
magnetic) current expressed in loop space terms: \cite{Chanftsou}
\begin{equation}
J_{\mu\nu}[\xi|s] = \tilde{g}\,\epsilon_{\mu\nu\rho\sigma}\dot\xi^\sigma(s)
[\bar\psi(\xi(s))\gamma^\rho T^i\psi(\xi(s))]\Omega^{-1}_\xi(s,0)\tau_i
\Omega_\xi(s,0),
\label{Jmunu}
\end{equation}
which is to be substituted into the topological constraint (\ref{Gausslaw})
defining the monopole charge at $\xi(s)$. Here, 
\begin{equation}
\Omega_\xi(s,0)=\omega(\xi(s_+))\Phi_\xi(s_+,0),
\label{Omegaxis0}
\end{equation}
where $\Phi_\xi(s,0)$ is the parallel phase transport from the reference 
point $P_0$ to the point $\xi(s)$, and $\omega(x)$ is a local transformation
matrix which rotates from the frame in which the field is measured to the 
frame in which the ``phase'' of the monopole is measured.  An important 
point here is the appearance of $s_+$ in the argument of 
$\Phi_\xi(s_+,0)$ which represents $s + \epsilon/2$ with $\epsilon > 0$ where
$\epsilon$ is taken to zero after the functional differentiation and
integration in $\xi$ have been performed. \cite{Chanftsou, Chanftsou1}
As a result, $\Omega_\xi(s,0)$ satisfies, for example,
\begin{equation}
\Omega_\xi^{-1}(s,0){\delta\over{\delta\xi^\mu(s)}}
\Omega_\xi(s,0)=-ig\,F_\mu[\xi|s].
\label{omegaomega-1}
\end{equation}

The occurrence in $J_{\mu\nu}[\xi|s]$ of the factors $\Omega_\xi(s,0)$ and 
its inverse, both depending on the point $\xi(s)$, will make the integrations
we have to do rather awkward.  For this reason, we prefer to recast the
whole problem in terms of a new set of rotated, ``hatted'' variables:
\begin{equation}
\widehat{F}_\mu[\xi|s] = \Phi[\xi] F_\mu[\xi|s] \Phi^{-1}[\xi],
\label{Fhat}
\end{equation}
and
\begin{equation}
\widehat{L}_{\mu\nu}[\xi|s] = \Phi[\xi] L_{\mu\nu}[\xi|s] \Phi^{-1}[\xi].
\label{Lhat}
\end{equation}
Note that these hatted variables no longer depend on the early part of $\xi$
from $s' = 0$ to $s' = s$ as the original variables $F_\mu[\xi|s]$ and
$L_{\mu\nu}[\xi|s]$ do, but rather on the later part of the loop with 
$s_-\leq s'\leq 2\pi$, where $s_- = s - \epsilon/2$, with $\epsilon$ being 
a positive infinitesimal quantity.  This can be seen by observing that
$F_\mu= {i\over g}\Phi^{-1}\delta_\mu \Phi$ and therefore $\widehat
F_\mu= {i\over g}(\delta_\mu\Phi)\Phi^{-1}$, which is a function of $\xi(s')$
with $s_-\leq s'\leq 2\pi$.  In terms of the hatted variables, we have for 
(\ref{calA}):
\begin{eqnarray}
{\widehat {\cal A}} & = & \int\!\delta\xi\,ds\,a_\xi\,Tr\,\left \{\widehat
   F_\mu\widehat F^\mu\right \} + \int\!d^4x\,\bar\psi\,(i\partial_\mu
   \gamma^\mu - m)\,\psi    \nonumber\\ 
   & & +\int\!\delta\xi\,ds\,Tr\,\left \{ \widehat L_{\mu\nu}\left
[ \delta^\nu\widehat F^\mu-\delta^\mu\widehat F^\nu-ig
\, \left [\widehat F^\mu,\widehat F^\nu\right ] + 4\pi \widehat
J^{\mu\nu}\right ]\right \},
\label{actionhat}
\end{eqnarray}
where $\widehat{J}_{\mu\nu}[\xi|s]$ differs from $J_{\mu\nu}[\xi|s]$ only by
having $\Omega_\xi(s,0)$ in (\ref{Jmunu}) replaced by 
\begin{equation}
\widehat\Omega_\xi(s,0) = \Omega_\xi(s,0) \Phi^{-1}[\xi],
\label{Omegahat}
\end{equation}
which is independent of the early part of the loop up to and including the 
point $\xi(s)$ since by (\ref{omegaomega-1}), for $s'<s_+$:
\begin{eqnarray}
\delta_\mu(s')\widehat\Omega_\xi(s,0)&=&\delta_\mu(s')\left [\omega(\xi(s_+))
\Phi_\xi(s_+,0)\Phi^{-1}[\xi]\right]\nonumber\\
&=&\omega(\xi(s_+))\delta_\mu(s')[\Phi_\xi(2\pi,s_+)]=0.
\label{omegahat2} 
\end{eqnarray}
As we shall see, this property of $\widehat\Omega_\xi$ in 
$\widehat J_{\mu\nu}[\xi|s]$ will make our task in evaluating Feynman 
integrals much easier. 

In terms of the hatted variables, the partition function $Z$ appears now 
as:\footnote{There can in principle be a Jacobian of transformation in the 
integral depending on which variable one chooses originally to quantize in, 
whether $\widehat F_\mu[\xi|s]$, $F_\mu[\xi|s]$ or $A_\mu(x)$.  However, 
working to order $g^0$ as we do here we need not bother, the Jacobian between 
any pair of these variables being then just a constant factor.  To higher 
orders, it will matter.  In fact our inability as yet to handle the Jacobian 
is one main reason preventing us from going to higher orders in $g$ at present.}
\begin{equation}
\widehat Z = \int\!\delta\widehat F\delta\widehat L \delta\psi \delta\bar\psi \,
   \exp{i\widehat {\cal A}}.
\label{Zhat}
\end{equation}
Since we shall be working exclusively with these hatted variables from now on, 
we shall henceforth drop the ``hat'' in our notation, assuming it now to be 
understood.  We shall also suppress the arguments of the field variables
unless this should lead to ambiguities.

We shall try now to evaluate the integral (\ref{Zhat}) to order $0$ in $g$
starting with the integral in $F$. To this order, ${\cal A}$ is quadratic
in $F$ so that the integral in $F$ is Gaussian and can be evaluated just by 
completing squares.  This brings about a term of the form $\delta_\alpha
L^{\mu\alpha} \delta^\rho L_{\mu\rho}$ in the exponent of the resulting
integrand which is thus again quadratic in the variable $L_{\mu\nu}$.  To
evaluate next the integral in $L$, we encounter a problem in completing 
the square for $L_{\mu\nu}$, due to the noninvertibility of the projection 
operator involved in the quadratic term.  This is a reflection of the fact
that in $L$ there is a gauge redundancy.  Although $L_{\mu\nu}$, like the
Polyakov variable $F_\mu$, is by construction gauge invariant (apart from
an unimportant $x$-independent gauge rotation at the reference point $P_0$)
under the original Yang-Mills gauge transformation, there is another gauge
symmetry of the theory \cite{Chanftsou} under which $L_{\mu\nu}$ transforms 
like an antisymmetric tensor potential of the Freedman-Townsend type 
\cite{Freedsend}.  Thus, in order to complete the square for $L_{\mu\nu}$ 
and integrate this field out, we need to impose a gauge-fixing condition 
on $L_{\mu\nu}$ by taking advantage of this new $\widetilde U$-symmetry of 
the theory.  

We propose then to impose on $L_{\mu\nu}$ the following gauge condition:
\cite{Faridani,Fosco}
\begin{equation}
C_\mu(L)=\epsilon_{\mu\nu\rho\sigma}\delta^\nu(s) L^{\rho\sigma}[\xi|s]=0.
\label{gaugecond} 
\end{equation}
In this gauge, which can be shown to be always possible \cite{Faridani}, the 
transverse degrees of freedom of $L_{\mu\nu}$ do not propagate and only the 
longitudinal ones are physical.  Following the standard procedures, we 
introduce then the suppression factor $C^2$ and the Faddeev-Popov determinant 
$\Delta_1$ as follows:
\begin{equation}
C^2=\int\!\delta\xi\,ds\,\,{1\over{2\alpha(s)}}\,
\epsilon_{\mu\nu\rho\sigma}\epsilon^{\mu\alpha\beta\gamma}\,Tr\,\left
[\delta^\nu L^{\rho\sigma}\delta_\alpha L_{\beta\gamma}\right ],
\end{equation}
and:
\begin{equation}
\Delta_1=\int\!\delta\eta\,\delta{\bar\eta}\,\exp i\int\!\delta\xi\,ds
   \,Tr\,\left \{ {\bar\eta}_\mu[\xi|s]\left ({{\delta C^{'\mu}(L^\Lambda)}
   \over{\delta\Lambda^\nu}}\right )\eta^\nu[\xi|s] \right \},
\end{equation}
where $\eta$ and ${\bar\eta}$ are two independent vector-valued Grassmann 
variables depending on the later part of the loop, and $C^{'\mu}(L^\Lambda)$ 
is obtained by applying to (\ref{gaugecond}) a $\widetilde U$-transformation
with gauge parameter $\Lambda_\beta[\xi|s]$ \cite{Chanftsou,Faridani}, thus:
\begin{equation}
C'_\mu(L^\Lambda) = C_\mu(L)+\epsilon_{\mu\nu\rho\sigma}\delta^\nu(s)
   \widetilde\Delta L^{\rho\sigma}[\xi|s],
\end{equation}
for ${\widetilde\Delta} L^{\rho\sigma} = \epsilon^{\rho\sigma\alpha\beta}
\delta_\alpha \Lambda_\beta$. The path-integral in (\ref{Zhat}) then becomes:  
\begin{eqnarray}
Z & = & \int\!\delta L\,\delta F\,\delta \eta\,\delta {\bar\eta}
   \,\delta \psi\,\delta {\bar\psi}\,\exp{i\int\!d^4x\,\bar\psi(i\partial_\mu
   \gamma^\mu - m)\psi}   \nonumber\\ 
  &   &\exp i\int\!\delta\xi\,ds\,Tr\,\left \{a_\xi\, F_\mu F^\mu+
   L_{\mu\nu}\left [\delta^\nu F^\mu-\delta^\mu F^\nu-ig\,
   [F^\mu, F^\nu] + 4\pi J^{\mu\nu}\right ]\right.\nonumber\\
  &   &\left.+(2\alpha)^{-1}\epsilon_{\mu\nu\rho\sigma}
   \epsilon^{\mu\alpha\beta\gamma}\,\delta^\nu L^{\rho\sigma}\,
   \delta_\alpha L_{\beta\gamma}+2{\bar\eta}_\mu(g^{\mu\nu}\Box_\xi
   -\delta^\mu\delta^\nu) \eta_\nu\right \},
\label{genfunc}
\end{eqnarray}
where $\Box_\xi$ denotes:
\begin{equation}
\Box_\xi(s) = \frac{\delta^2}{\delta\xi^\mu(s) \delta\xi_\mu(s)}.
\end{equation}
Here, $\alpha(s)=2a_\xi(s)$ is chosen such that the gauge-fixing term for 
$L_{\mu\nu}$ cancels the term $\delta_\alpha L^{\mu\alpha}\delta^\rho
L_{\mu\rho}$ brought about by completing the square for $F_\mu$.

In (\ref{genfunc}), we see again the appearance of a non-invertible operator
$g^{\mu\nu}\Box_\xi-\delta^\mu\delta^\nu$.  In order to integrate out the
fields $\eta$ and ${\bar\eta}$ and eliminate the off-diagonal term 
${\bar\eta}_\mu\delta^\mu\delta^\nu\eta_\nu$, we must fix the gauge for a 
second time, by finding a second gauge-symmetry of the action.  This is 
accomplished by writing the last term in (\ref{genfunc}) as:
\begin{equation}
\int\!\delta\xi\,ds Tr\left [{\bar\eta}_\mu(g^{\mu\nu}\delta_\rho\delta^\rho
-\delta^\mu\delta^\nu) \eta_\nu\right ]=-\int\!\delta\xi\,ds Tr\left
[(\delta_\nu {\bar\eta}_\mu-\delta_\mu {\bar\eta}_\nu)
(\delta^\nu \eta^\mu-\delta^\mu \eta^\nu)\right ],
\end{equation}
which we notice is invariant under:
\begin{equation}
\eta^\mu[\xi|s] \to \eta^\mu[\xi|s]+\delta^\mu \lambda[\xi|s].
\end{equation}
This symmetry allows us to ``fix the gauge'' for $\eta$ by choosing $\lambda$ 
such that:
\begin{equation}
\delta_\mu \eta^{'\mu}=\delta_\mu \eta^\mu+\Box_\xi \lambda=0,
\end{equation}
which is always possible, given initial conditions for $\lambda$.

Including the suppression factor:
\begin{equation}
D^2=\int\!\delta\xi\,ds\,{1\over{2\beta}}\,Tr\,\left
[(\delta_\mu {\bar\eta}^\mu)(\delta_\nu \eta^\nu)\right ],
\end{equation}
as well as the Faddeev-Popov determinant:
\begin{equation}
\Delta_2=\int\!\prod_{i=1,2}\,\delta \phi_i \,\delta {\bar\phi_i}\,
   \exp i\int\!\delta\xi\,ds\,{1\over{2\beta}}\,Tr\, \sum_{i=1,2}\left 
   \{ {\bar\phi}_i \Box_\xi \phi_i \right \},
\end{equation}
in $(\ref{genfunc})$ yields:
\begin{eqnarray}
Z & = & \int\!\delta L\,\delta F\,\delta \eta\,\delta {\bar\eta}\,
   \delta\psi\delta\bar\psi \prod_{i=1,2}\,\delta \phi_i\,\delta {\bar\phi}_i\,
   \exp{i\int\!d^4x\,\bar\psi(i\partial_\mu\gamma^\mu - m)\psi}\nonumber\\
  &   & \exp i\int\!\delta\xi\,ds\,Tr\,\left \{a_\xi\, F_\mu F^\mu
   +2\, {\bar\eta}_\mu\Box_\xi \eta^\mu\right.\nonumber\\
  &   & + L_{\mu\nu}\left [\delta^\nu F^\mu-\delta^\mu F^\nu 
   -ig\,[F^\mu, F^\nu] + 4\pi J^{\mu\nu}\right ]
   -{1\over 4} \sum_{i=1,2}{\bar\phi}_i\,\Box_\xi \phi_i \nonumber\\
  &   & \left.\qquad\qquad\qquad\qquad\qquad+(4a_\xi)^{-1}
   \epsilon_{\mu\nu\rho\sigma} \epsilon^{\mu\alpha\beta\gamma}
   \delta^\nu L^{\rho\sigma} \delta_\alpha L_{\beta\gamma}\right \}.
\end{eqnarray}
where $\phi_i[\xi|s]$ and ${\bar\phi}_i[\xi|s]$ for $i=1,\,2,$ are four 
independent commuting fields depending on the later part of the loop 
$\xi$ \cite{Namazie,Siegel}, and we have chosen $\beta=-1/2$, in order for 
the second gauge-fixing condition to cancel the off-diagonal term in 
${\bar\eta}_\mu\delta^\mu\delta^\nu\eta_\nu$.

The ghosts $\eta$, ${\bar\eta}$, $\phi_i$ and ${\bar\phi}_i$ can easily be 
integrated out and decouple from the theory. The effective action after 
integrating out the ghost fields is:
\begin{eqnarray}
{\cal A}_{eff} & = & \int\!\delta\xi\,ds\,Tr\,\left \{a_\xi\, F_\mu F^\mu
   + 2\,\bar\psi(i\partial_\mu\gamma^\mu - m)\psi\right.\nonumber\\
  &   & + L_{\mu\nu}\left [\delta^\nu F^\mu-\delta^\mu F^\nu 
   -ig\,[F^\mu, F^\nu] + 4\pi J^{\mu\nu}\right ]  \nonumber\\
  &   & \left.\qquad\qquad\qquad\qquad\qquad+(4a_\xi)^{-1}
   \epsilon_{\mu\nu\rho\sigma} \epsilon^{\mu\alpha\beta\gamma}
   \delta^\nu L^{\rho\sigma} \delta_\alpha L_{\beta\gamma}\right \}.
\label{actionhat2}
\end{eqnarray}
The decoupling of the ghosts can be explained as follows: although the theory
itself is nonabelian in character, the $\widetilde U$-transformation on 
$L_{\mu\nu}$ does not involve a term coupling $\Lambda_\mu$ to $L_{\mu\nu}$
\cite{Chanftsou}, in contrast to the usual Yang-Mills ``$U$''-transformation 
on the gauge potential $A_\mu(x)$.  In the following, we shall assume 
that the ghost fields have all been integrated out, and drop henceforth the
subscript $eff$ to ${\cal A}$ from our notation.

\setcounter{equation}{0}
\section{Generating Functional, Propagators, and Vertex} 

To manage the perturbation expansion we shall adopt the usual generating 
functional method.  One starts by considering in the action only the ``free 
field'' terms of order $2$ and lower in the fields, denoted generically say by 
$H$, and ignoring all ``interaction'' terms of higher order.  External 
current terms are then added to this action.  On integrating out the 
fields $H$ by completing squares one obtains the free-field generating 
functional $Z^{(2)}[{\cal J}]$ which depends on the external current ${\cal J}$ 
only.  The propagator for any field $H$ say, will then be given by the 
expression:
\begin{equation}
\langle H(x)H(y)\rangle
={1\over i}{\delta\over {\delta {\cal J}(x)}}{1\over i}{{\delta}\over
{\delta {\cal J}(y)}}\,Z^{(2)}[{\cal J}]|_{{\cal J}=0},
\label{functderiv}
\end{equation}
where ${\cal J}$ here denotes the external current which corresponds to $H$.
Next, collecting the higher order ``interaction'' terms of the action, say, 
${\cal A}_I[H]$, one can write the full generating functional formally as:
\begin{equation}
Z[{\cal J}] = \exp\left[i{\cal A}_I[-i\delta/\delta{\cal J}]\, \right]\, 
   Z^{(2)}[{\cal J}].
\label{eqgenf}
\end{equation} 
Any term in the perturbation expansion can then be obtained by taking the
appropriate derivative of $Z[{\cal J}]$ with respect to ${\cal J}$.

In our problem here formulated in loop space, the terms of the action 
${\cal A}$ in (\ref{calA}) which are second order or lower in the fields 
$\psi$, $F_\mu$ and $L_{\mu\nu}$ do not correspond to just the free action 
${\cal A}^0$ of (\ref{calA0}), and the concept of interaction has been 
replaced by that of a constraint imposed through the Wu-Yang criterion.  
Nevertheless, the method can still be applied.  Writing then the action 
(\ref{calA}) to second order in the fields, including external current terms, 
we have:
\begin{eqnarray}
{\cal A}^{(2)}[{\cal J}] & = & \int\!\delta\xi\,ds\,\left \{a_\xi F_\mu^i 
   F^\mu_i + L_{\mu\nu}^i\left (\delta^\nu F^\mu_i - \delta^\mu F^\nu_i\right )
   + {\cal J}^{\mu\nu}_i L^i_{\mu\nu} + {\cal J}_\mu^i F^\mu_i\right.\nonumber\\
  &   & \left. + (2\alpha)^{-1}\epsilon_{\mu\nu\rho\sigma}\,
   \epsilon^{\mu\alpha\beta\gamma} \delta^\nu L_i^{\rho\sigma}
   \delta_\alpha L_{\beta\gamma}^i\right \} + \int\!d^4x\,
   \left [(\bar{\cal J}\psi+\bar\psi{\cal J}) + \bar\psi(i\partial_\mu\gamma^\mu
   - m)\psi\right ]    \nonumber\\
  & = & {\cal A}^{(2)}_F + {\cal A}^{(2)}_M.
\label{action2}
\end{eqnarray}
The generating functional factors into a gauge term and a matter term where
the matter term is the same as in ordinary local formulations.  For the gauge 
term written in loop space:
\begin{equation}
Z^{(2)}_{F} = \int\!\delta L \delta F \exp i{\cal{A}}^{(2)}_F,
\end{equation}
after completing the squares for $F_\mu$ and $L_{\mu\nu}$ and integrating 
them out, we obtain, up to a multiplicative factor:
\begin{eqnarray}
Z^{(2)}_{F}[{\cal J}] & = & \exp -i\int\delta\xi ds\frac{1}{4a_\xi}\left
   [2\Box_{\xi}^{-1} (\delta^{\nu}{\cal J}^{\mu}_{i} -a_{\xi} 
   {\cal J}^{\mu\nu}_{i})^{2} + {\cal J}^{i}_{\mu}{\cal J}_{i}^{\mu}\right]
   \nonumber\\
  & = & \exp -i\int\delta\xi ds\Big{[}\frac{1}{2a_\xi}\Box_{\xi}^{-1}
   \delta^{\nu}{\cal J}^{\mu}_{i}\delta_{\nu}{\cal J}^{i}_{\mu}-\Box_{\xi}^{-1}
   \delta^{\nu}{\cal J}^{\mu}_{i}{\cal J}_{\mu\nu}^{i}\nonumber\\
  &   &\qquad
   \; \; \; \; \; \; \; \; \; \; \; \; \; \; \; \; \; \; \; \; \; \; \; \; \; 
   + \frac{a_\xi}{2}\Box_{\xi}^{-1}{\cal J}^{\mu\nu}_{i}{\cal J}^{i}_{\mu\nu}
   + \frac{1}{4a_{\xi}}{\cal J}^{i}_{\mu}{\cal J}_{i}^{\mu}\Big{]}.
\label{freefgenf}
\end{eqnarray}
Differentiating functionally with respect to the currents yields for the
propagators:
\begin{equation}
\langle F_\mu^i[\xi|s] F_{\mu'}^{i'}[\xi'|s']\rangle=-{3i\over{4a_{\xi'}(s')}}
   \delta^{ii'} g_{\mu\mu'}\,\delta(s-s')\prod_{\bar s = 0}^{2\pi}\,
   \delta^4(\xi(\bar s)-\xi'(\bar s)),
\label{Fprop}
\end{equation}
and:
\begin{equation}
\langle L^i_{\mu\nu}[\xi|s] L^{i'}_{\mu'\nu'}[\xi'|s'] \rangle
   = {{i\,a_{\xi'}(s')} \over 2}\delta^{ii'}\delta (s-s')\,\Box_{\xi'}^{-1}(s')
   \prod_{\bar s = 0}^{2\pi}\delta^4(\xi(\bar s)-\xi'(\bar s))\,
   (g_{\mu\mu'}g_{\nu\nu'}-g_{\mu\nu'}g_{\nu\mu'}).
\label{Lprop}
\end{equation}
Note that for the functional differentiation above, we have used:
\begin{eqnarray}
{{\delta X_\mu^i[\xi|s]}\over{\delta X^\alpha_j[\xi'|s']}}
  & = & \delta^{ij}\,g_{\mu\alpha}\,\delta(s-s')\,\prod_{\bar s=0}^{2\pi}\,
   \delta^4(\xi(\bar s)-\xi'(\bar s)),\nonumber\\
   {{\delta Y^i_{\mu\nu}[\xi|s]}\over {\delta Y_j^{\alpha\beta}[\xi'|s']}}
  & = & {1\over 2}\delta^{ij}(g_{\mu\alpha}g_{\nu\beta}
   -g_{\mu\beta}g_{\nu\alpha})\,\delta(s-s')\,
   \prod_{\bar s=0}^{2\pi}\,\delta^4(\xi(\bar s)-\xi'(\bar s)).
\end{eqnarray}

To order $g^0$, the commutator term in the loop space curvature $G_{\mu\nu}$ 
can be dropped so that there remains only one term in the action ${\cal A}$
of (\ref{calA}) which is of higher than second order, namely the term coming
from the current $J_{\mu\nu}$ in the constraint:
\begin{equation}
{\cal A}^{(3)} = 4\pi\widetilde{g} \int\delta\xi ds\epsilon_{\mu\nu\sigma\rho}
   \, \dot{\xi}^\sigma(s)\,\bar{\psi}(\xi(s)) \Omega_{\xi}(s,0)T^{i}\,
   \Omega_\xi^{-1}(s,0)\, \psi(\xi(s))\gamma^\rho L^{\mu\nu}_{i}[\xi|s]\,.
\label{calA3}
\end{equation}
This can be substituted as ${\cal A}_I$ in (\ref{eqgenf}) to construct the
generating functional $Z[{\cal J}]$.  Since the ``interaction'' only involves 
the fields $\psi$ and $L_{\mu\nu}$ and not $F_\mu$, we can put ${\cal J}_\mu^i$,
the external current for $F_\mu$, equal to zero in (\ref{freefgenf}), keeping
only the remaining relevant terms.  If we denote the propagator of $\psi$ by
$S_{F}(x-y)$ and the propagator (\ref{Lprop}) for $L_{\mu\nu}$ by 
$\Delta_{\mu\nu,\mu'\nu'}[\xi,\xi'|s,s']$, the free field generating 
functional up to factors can then be written as:
\begin{eqnarray}
Z^{(2)}[{\cal J}] & = & \exp -i\int d^{4}xd^{4}y\, \bar{{\cal J}}(x)
   S_{F}(x-y){\cal J}(y) \nonumber\\
   &   & \exp -\frac{i}{2}\int \delta\xi\delta\xi'dsds' 
   {\cal J}_j^{\mu\nu}[\xi|s] \Delta^{jj'}_{\mu\nu,\mu'\nu'}[\xi,\xi'|s,s']
   {\cal J}_{j'}^{\mu'\nu'}[\xi'|s'].
\label{Z2calJ}
\end{eqnarray}  
Applying the operation as indicated in (\ref{eqgenf}) to this will give us
the full generating functional we want.

For example, suppose we are interested in the ``interaction vertex'', we 
expand (\ref{eqgenf}) to first order in $\widetilde{g}$, obtaining after
a straightforward calculation, up to numerical factors:
\begin{eqnarray}
Z[{\cal J}] &= & \Bigg[1+ 4\pi i\widetilde{g}\int\!\delta\xi ds\, 
   \epsilon_{\mu\nu\sigma\rho} \Omega_\xi^{kl}(s,0)T^{i}_{lm}\,
   \Omega_\xi^{-1\, mn}(s,0)\, \gamma^\rho\, \dot{\xi}^\sigma(s) \nonumber \\ 
   &   &  +\, \int d^{4}y\, S_{F}^{n\ell}(\xi(s)-y)\, 
   {\cal J}_{\ell}(y)\int d^{4}x\, \bar{{\cal J}}_{j}(x)S_{F}^{jk}(x-\xi(s))
   \nonumber \\
   &   & \qquad\qquad\int\delta\xi'ds'\, 
   \Delta^{ii'}_{\mu\nu,\mu'\nu'}[\xi,\xi'|s,s']\, 
   {\cal J}^{\mu'\nu'}_{i'}[\xi'|s']\,\Bigg]\, Z^{(2)}[{\cal J}],
\label{eqgenfirst}
\end{eqnarray}
where we have dropped the vacuum term $S_F(0)$ and, 
to avoid confusion, we have written out explicitly the internal symmetry 
indices.  Differentiating with respect to the appropriate currents then
yields:
\begin{eqnarray}
& &(-i)^{3}\,
\frac{\delta}{\delta{\cal J}^{l}_{\mu\nu}[\xi'|s']}\frac{\delta}
{\delta{\cal J}^{m}(x_{2})}\frac{\delta}{\delta\bar{{\cal J}}^{i}(x_{1})}\,
Z[{\cal J}]\Big |_{{\cal J}=0}\nonumber\\
& = & - 4 \pi \widetilde{g} \int \delta \xi ds \,
S_{F}^{ik}(x_{1}-\xi(s)) \Omega_\xi^{kl'}(s,0)T^j_{l'm'}
\Omega_\xi^{-1\ m'n}(s,0) S_{F}^{nm}(\xi(s)-x_{2}) \nonumber \\
&   & \;\;\;\;\;\;\; \epsilon^{\mu'\nu'\rho\sigma} \gamma_{\rho} 
\dot{\xi}_\sigma(s) \Delta^{lj}_{\mu\nu,\mu'\nu'}[\xi,\xi'|s,s'],
\label{vertex}
\end{eqnarray}
where the vertex is obtained by eliminating the propagators from the
external lines.

\setcounter{equation}{0}
\section{The dual potential ${\tilde A}_\mu(x)$}

Before we proceed to work out explicitly the loop space formulae for the
interaction vertex and generating functional, we shall first introduce a
quantity ${\tilde A}_\mu(x)$ found in an earlier paper \cite{Chanftsou} 
which will considerably simplify our task:
\begin{equation}
\widetilde{A}_{\mu}(x)=4\pi\int \delta\xi ds\, \epsilon_{\mu\nu\rho\sigma}
{\Omega}_{\xi}(s,0)T^{i}{\Omega}_{\xi}^{-1}(s,0)
L^{\rho\sigma}_{i}[\xi|s]\dot{\xi}^{\nu}(s)\delta^{4}(x-\xi(s)).
\label{Atilde}
\end{equation}
In the classical theory, it is now known \cite{Chanftsou1} that this quantity
${\tilde A}_\mu(x)$ plays an exactly dual role to the ordinary Yang-Mills
potential $A_\mu(x)$, acting as the parallel phase transport for the monopole
wave function and giving a complete description of the dual field.  This last
statement by itself does not necessarily imply that ${\tilde A}_\mu(x)$ will 
play the same role also in the quantum field theory but, as we shall see, 
it turns 
out to do so, at least to order $g^0$. 

We note first that the gauge fixing condition that we have imposed on the
loop variable $L_{\mu\nu}$ is in fact equivalent to the standard Lorentz 
condition on the local quantity ${\tilde A}_\mu(x)$.  This can be seen as
follows.  Differentiating ${\tilde A}_\mu(x)$ in (\ref{Atilde}) with respect 
to $x$ and then integrating by parts with respect to $\xi$ with the help 
of $\delta^4(x-\xi(s))$, we obtain:
\begin{equation}
\partial^{\mu}\widetilde{A}_{\mu}(x) = -4\pi\int\delta\xi ds\, 
   \epsilon_{\mu\nu\rho\sigma}\, \delta^{\mu}(s)\left[\Omega_\xi(s,0)
   L^{\rho\sigma}[\xi|s] \Omega_{\xi}^{-1}(s,0)\right] \dot{\xi}^\nu(s)
   \delta^4(x-\xi(s)).
\end{equation}
Recall now the fact that we are working with what we called ``hatted''
variables so that by (\ref{omegahat2}) the derivative $\delta^\mu(s)$ above
commutes with $\Omega_\xi(s,0)$ and acts only on $L^{\rho\sigma}[\xi|s]$.
We see then that the gauge-fixing condition (\ref{gaugecond}) that we
have imposed on $L_{\mu\nu}[\xi|s]$ is indeed equivalent to the Lorentz 
condition:
\begin{equation}
\partial^{\mu}\widetilde{A}_{\mu}(x)=0.
\label{duallor}
\end{equation}

Secondly, we note that to order $g^0$ and in the absence of its interactions 
with $\psi$, $\widetilde{A}^{\mu}(x)$ satisfies the free field equation:
\begin{equation}
\Box\widetilde{A}_{\mu}(x)=0,
\label{boxatilde}
\end{equation}
which allows for its expansion into the usual plane wave creation/annihilation 
operators.  The loop-space curvature $G_{\mu\nu}=0$ in the absence of the
interaction term.  To zeroth order in $g$, the curvature is given by
$G_{\mu\nu}=\delta_{\mu}F_{\nu}-\delta_{\nu}F_{\mu}$. The equation 
$F_{\mu}=a_{\xi}^{-1}\, \delta^{\nu}L_{\mu\nu}$ also holds, since to zeroth 
order in $g$, the covariant loop space derivative is the same as the ordinary 
loop space derivative. For the same reason, the gauge fixing condition implies 
$\epsilon_{\mu\nu\rho\sigma}\, \delta^{\nu}L^{\rho\sigma}=0$. Inserting 
$F_{\mu}$ and using this expression we obtain:
\begin{equation}
\Box_{\xi}L_{\mu\nu}=0.
\label{eqloopbox}
\end{equation}
On the other hand:
\begin{equation}
\Box\widetilde{A}_{\mu}(x)=~4\pi\int\delta\xi
ds\, \epsilon_{\mu\nu\rho\sigma}\Box_{\xi}\left[\Omega_{\xi}(s,0)
L^{\rho\sigma}[\xi|s]\Omega^{-1}_{\xi}(s,0)\right]\dot{\xi}^{\nu}(s)\,
\delta^{4}(x-\xi(s)).
\end{equation}
For the same reasons as before, the derivatives could be taken inside the 
bracket to act on $L_{\rho\sigma}$ only. Using (\ref{eqloopbox}) we then 
obtain (\ref{boxatilde}) as required.

Finally, we show that (\ref{eqgenf}) can in fact be expressed in terms 
of the dual potential $\widetilde{A}_{\mu}(x)$ and a corresponding local 
current $\widetilde{j}^{\mu}(x)$ instead of the $L$-field and its 
corresponding current in loop space.  We note first that the ``interaction''
${\cal A}_I$ in (\ref{eqgenf}) or, in other words, ${\cal A}^{(3)}$ of 
(\ref{calA3}), can be rewritten as:
\begin{equation}
{\cal A}^{(3)} = \widetilde{g} \int d^4 x
\bar{\psi}(x) \widetilde{A}_\mu(x) \gamma^\mu   \psi(x),
\end{equation}
for $\widetilde{A}_\mu(x)$ as given in (\ref{Atilde}), so that it is a function 
of $L_{\mu\nu}$ only in that particular combination.  We need therefore 
introduce a current really only for this combination of $L_{\mu\nu}$, namely a 
local current $\widetilde{j}^{\mu}(x)$ corresponding to $\widetilde{A}_\mu(x)$,
by incorporating in the action a term of the form:
\begin{equation}
\int d^4x \widetilde{A}_i^\mu(x) \widetilde{j}^i_\mu(x).
\label{Ajtilde}
\end{equation}
This can be rewritten in the original form given in (\ref{action2}):
\begin {equation}
\int \delta\xi ds\,L_{\rho\sigma}^{i}[\xi|s]\, {\cal J}^{\rho\sigma}_{i}[\xi|s],
\end{equation}
provided that ${\cal J}^{\mu\nu}_i[\xi|s]$ is of the special form:
\begin{equation}
{\cal J}^{\mu\nu}_{i}[\xi|s] = 4\pi\int d^4x\, \epsilon^{\mu\nu\rho\sigma}
   \left[\Omega_{\xi}(s,0)T_{i}\Omega_{\xi}^{-1}(s,0)\right]^{j}
   \dot{\xi}_{\sigma}(s)\widetilde{j}_{\rho}^{j}(x)\delta^{4}(x-\xi(s)).
\label{eqcurr}
\end{equation}
Substituting this ${\cal J}^{\mu\nu}_i[\xi|s]$ into $Z^{(2)}[{\cal J}]$ of 
(\ref{Z2calJ}) and $Z[{\cal J}]$ of (\ref{eqgenfirst}), one easily obtains that
up to numerical factors:
\begin{equation}
Z^{(2)}[{\cal J}] = \exp-\frac{i}{2} \int d^{4}x\, d^{4}y\, 
   \left[\bar{{\cal J}}(x)S_{F}(x-y){\cal J}(y)+2\widetilde{j}_j^{\mu}(x)\, 
   \left\langle\widetilde{A}^{j}_{\mu}(x)\widetilde{A}^{j'}_{\mu'}(y)
   \right\rangle\, \widetilde{j}_{j'}^{\mu'}(y)\, \right],
\label{eqgenfunc}
\end{equation}\\
and that:
\begin{eqnarray}
Z[{\cal J}]& = &\Bigg[1+i\widetilde{g}\int d^{4}x\, d^{4}y\, d^{4}z\, 
d^{4}w\, \bar{{\cal J}}^{i}(x)S_{F}^{ik}(x-w)T_{j}^{kn}\, \nonumber\\
& &S_{F}^{ni'}(w-y){\cal J}^{i'}(y)\gamma^{\rho}
\left\langle\widetilde{A}^{j}_{\rho}(x)\widetilde{A}^{j'}_{\rho'}(z)
\right\rangle\,
\widetilde{j}^{\rho'}_{j'}(z)\, \Bigg]\, Z^{(2)}[{\cal J}],
\label{ZJinAtilde}
\end{eqnarray}
with:
\begin{eqnarray}
\langle\widetilde{A}^{i}_{\mu}(x)\widetilde{A}^{i'}_{\mu'}(x')\rangle &=& 
16\pi^{2}
\int\delta\xi\delta\xi'
dsds'\epsilon_{\mu\nu\rho\sigma}\epsilon_{\mu'\nu'\rho'\sigma'}
\dot{\xi}^{\nu}(s)\dot{\xi}'^{\nu'}(s')\nonumber\\
& &\left[\Omega_{\xi}(s,0)T^{j}\,
\Omega_{\xi}^{-1}(s,0)\right]^{i}\left[\Omega_{\xi'}(s',0)T^{j'}\,
\Omega_{\xi'}^{-1}(s',0)\right]^{i'}\nonumber\\
& &\Delta_{jj'}^{\rho\sigma,\rho'\sigma'}[\xi,\xi'|s,s']\, 
\delta^{4}(x-\xi(s))\delta^{4}(x'-\xi'(s')).
\label{Atildeprop}
\end{eqnarray}

The formulae (\ref{eqgenfunc}) and (\ref{ZJinAtilde}) are formally the same
as those in standard Yang-Mills theory.  Indeed, if the quantity 
$\langle\widetilde{A}^i_\mu(x) \widetilde{A}^{i'}_{\mu'}(x')\rangle$ can be
identified with the standard propagator of the gauge potential in Yang-Mills
theory, then, apart from the gauge-boson self-interaction which has been
dropped in working only to order $g^0$, one would obtain exactly the same 
perturbation series in ${\tilde g}$ here as one does in $g$ in ordinary 
Yang-Mills theory.  At present, however, $\langle \widetilde{A}^i_\mu(x) 
\widetilde{A}^{i'}_{\mu'} (x') \rangle$ is still given in (\ref{Atildeprop}) 
as a complicated integral in loop space.  That this integral is in fact the 
same as the standard propagator of the Yang-Mills potential is the subject 
of the next section.

\setcounter{equation}{0}
\section{Loop Space Fourier Transform and the Propagator for ${\tilde A}_\mu$}

Inserting the expression for the $L$-field propagator from (\ref{Lprop}) into
(\ref{Atildeprop}), we have:
\begin{eqnarray}
\langle\widetilde{A}^{i}_{\mu}(x)\widetilde{A}^{i'}_{\mu'}(x')\rangle
T^{i}T^{i'}\!\!&\!\!=\!\!&\!\!8\,\pi^{2}\int \delta\xi\delta\xi'
ds\,\epsilon_{\mu\nu\rho\sigma}\epsilon_{\mu'\nu'\rho'\sigma'}a_{\xi'}(s)
\delta^{jj'}\dot{\xi}^{\nu}(s)\dot{\xi}'^{\nu'}(s)\nonumber\\
\!\!&\!\!&\!\!\left[\Omega_{\xi}(s,0)T^{j}\,
\Omega_{\xi}^{-1}(s,0)\right](g^{\rho\rho'}g^{\sigma\sigma'}-
g^{\rho\sigma'}g^{\sigma\rho'})
\Box_{\xi'}^{-1}(s)\prod_{\bar s=0}^{2\pi}\delta^4(\xi(\bar s)-\xi'(\bar s))
\nonumber\\
\!\!&\!\!&\!\!
\left[\Omega_{\xi'}(s,0)T^{j'}\Omega_{\xi'}^{-1}(s,0)\right]
\delta^{4}(x-\xi(s))\delta^{4}(x'-\xi'(s)),
\end{eqnarray}
where we have performed the $s'$ integration already.  This simplifies to:
\begin{eqnarray}
\langle\widetilde{A}^{i}_{\mu}(x)\widetilde{A}^{i'}_{\mu'}(x')\rangle
T^{i}T^{i'}&=&-16\pi^{2}\int \delta\xi\delta\xi'ds\, a_{\xi'}(s)
\dot\xi^\nu(s)\,\dot\xi'^{\nu'}(s)\left[\Omega_{\xi}(s,0)T^{j}\,
\Omega_{\xi}^{-1}(s,0)\right]\nonumber\\
& &G_{\mu\nu\mu'\nu'}\Box_{\xi'}^{-1}(s)\,\prod_{\bar s=0}^{2\pi}\,
\delta^4(\xi(\bar s)-\xi'(s))\left[\Omega_{\xi'}(s,0)T^{j'}\,
\Omega_{\xi'}^{-1}(s,0)\right]\nonumber\\
& &\qquad\qquad\qquad\qquad
\delta^{jj'}\delta^{4}(x-\xi(s))\delta^{4}(x'-\xi'(s)),
\end{eqnarray}
where we have used the abbreviation:
\begin{equation}
 G_{\mu\nu\mu'\nu'}=(g_{\mu\mu'}g_{\nu\nu'}-g_{\mu\nu'}g_{\nu\mu'}).
\end{equation}
Using the bra-ket notation of Dirac, we now define:
\begin{eqnarray}
\langle x|\Gamma^{\nu}_{j}|\xi\rangle     &=& 4\pi
i\,\dot\xi^\nu(s)\,\left[\Omega_{\xi}(s,0)T^{j}\, 
\Omega_{\xi}^{-1}(s,0)\right]\delta^{4}(x-\xi(s)),\label{eq1} \\
\langle\xi |\Delta_{\mu\nu\mu'\nu'}^{jj'} |\xi'\rangle &=& 
a_{\xi'}(s) G_{\mu\nu\mu'\nu'}
\delta^{jj'}\Box_{\xi'}^{-1}(s)\,\prod_{\bar
s=0}^{2\pi}\delta^4(\xi(\bar s)-\xi'(\bar s)),\label{eq2}  
%\langle\xi' |\Gamma'^{\nu'}_{j'}|x'\rangle &=& 4\pi i\, a_{\xi'}(s)
%\dot{\xi}^{\nu'}(s)\,\left[\Omega_{\xi'}(s,0)T^{j'}\, 
%\Omega_{\xi'}^{-1}(s,0)\right]\delta^{4}(x'-\xi'(s)), \label{eq3} \\
\end{eqnarray}
and write:
\begin{equation}
\langle\widetilde{A}^{i}_{\mu}(x)\widetilde{A}^{i'}_{\mu'}(x')\rangle
T^{i}T^{i'}=\int ds\,\langle x|\Omega_{\mu\mu'}|x'\rangle,
\end{equation}
with:
\begin{equation}
\Omega_{\mu\mu'} = \Gamma^{\nu}_{j}\Delta_{\mu\nu\mu'\nu'}^{jj'}
   \Gamma^{\nu'}_{j'}.
\label{Omegamumu'}
\end{equation}

Our aim now is to transform this propagator to momentum space so as to compare
with the standard propagator of the Yang-Mills potential.  Although this
propagator is itself an ordinary space-time quantity for which the Fourier 
transform is well-defined, it is expressed in terms of loop quantities the
Fourier transformation of which will require some care.  First, if in analogy 
to $\langle x|p\rangle=\exp{(-ipx)}$ in ordinary space-time, we define in 
loop space:
\begin{equation}
\langle\xi |\pi\rangle=\int dt\exp{i\,\xi(t)\pi(t)}.
\end{equation}
then we can write:
\begin{eqnarray}
\langle p|\Gamma^{\nu}_{j}|\pi\rangle&=&i\int d^{4}x\,
\delta\xi\, \langle p|x\rangle\langle
x|\Gamma|\xi\rangle\langle\xi|\pi\rangle\nonumber\\&=&i\int\delta\xi
\,e^{ip\,\xi(s)}\, 4\pi\dot\xi^\nu(s)\,\left[\Omega_{\xi}T^{j}\, 
\Omega_{\xi}^{-1}\right]\exp -i\int dt\, \xi(t)\pi(t),
\end{eqnarray}
\begin{equation}
\langle\pi'|\Gamma'^{\nu'}_{j}|p'\rangle=i\int\delta\xi'\,e^{-ip'\xi'(s)}4
\pi \dot{\xi}^{\nu'}(s)\left[\Omega_{\xi'}T^{j'}\Omega_{\xi'}^{-1}
\right]\exp
i\int dt\, \xi'(t)\pi'(t).
\end{equation}
and:
\begin{eqnarray}
\langle\pi|\Delta_{\mu\mu'\nu\nu'}^{jj'}|\pi'\rangle&=&\int\!\delta\xi\,
\delta\xi'\,a_\xi(s) \left
[ G_{\mu\nu\mu'\nu'}\delta^{jj'}\Box^{-1}_{\xi'}(s)\,\prod_{\bar
s=0}^{2\pi}\delta^4(\xi(\bar s)-\xi'(\bar s))\right ]\nonumber\\
& &\, \, \left(\exp i\int^{2\pi}_{0}dt\, \pi(t)\xi(t)\right)\, 
\left(\exp
-i\int_{0}^{2\pi}dt\, \pi'(t)\xi'(t)\right).
\label{piDeltapi'}
\end{eqnarray}
However, if we proceed now to evaluate these quantities, we shall find
$\delta$-functional ambiguities connected with the definition of the loop
derivative $\delta_\mu(s) = \delta/\delta \xi^\mu(s)$ and the tangent to
the loop $\dot{\xi}^\mu(s)$ both of which occur in the formulae above.  In 
other words, some regularization procedure is required in order to give 
these quantities an unambiguous meaning.  Our procedure, which we have
followed throughout our program \cite{Chanstsou,Chanftsou,Chanftsou1}, is
to replace first the $\delta$-function $\delta(s-s')$ inherent in the 
definition of the loop derivative by a bump function $\beta_\epsilon(s-s')$
of width $\epsilon$ and the tangent $\dot{\xi}^\mu(s)$ to the loop at $s$
by:
\begin{equation}
\dot{\xi}^{\mu}(s)=\frac{\xi^{\mu}(s_{+})-\xi^{\mu}(s_{-})}{s_{+}-s_{-}},
\label{xidotappr}
\end{equation}
for $s_{\pm}=s \pm \epsilon/2$, and then take the limit $\epsilon \rightarrow 0$
after the required operations have been performed.  We propose to follow the
same procedure here.

With these provisos we return to the evaluation of (\ref{piDeltapi'}).  The
box-operator there acts on the $\delta$-function, but by integrating by parts,
it can be made to act on the exponential function and the above expression 
becomes:
\begin{eqnarray}
\langle\pi|\Delta^{jj'}_{\mu\mu'\nu\nu'}|\pi'\rangle\!\!&=&\!\!
\int\!\delta\xi\,\delta\xi'\,
  G_{\mu\nu\mu'\nu'}\,\delta^{jj'}\prod_{\bar s=0}^{2\pi}\,
  \delta^4(\xi(\bar s)-\xi'(\bar s)) \,\exp
  i\int_0^{2\pi}dt\,\pi(t)\xi(t) \nonumber\\
&&\left \{\Box_{\xi'}^{-1}(s) \exp
-i\int_0^{2\pi}dt\,\pi'(t)\,\xi'(t)\right \}.
\end{eqnarray}
Simplifying further we obtain:
\begin{equation}
\langle\pi|\Delta^{jj'}_{\mu\mu'\nu\nu'}|\pi'\rangle=
\,  G_{\mu\nu\mu'\nu'}\prod_{\bar s=0}^{2\pi}\,\delta^4(\pi(\bar s)-
\pi'(\bar s))\int_{s_-}^{s_+}\!ds'\,ds''\,
{{\beta_\epsilon(s'-s)\beta_\epsilon(s''-s)}\over{\pi'_\alpha(s')
\pi^{'\alpha}(s'')}},
\label{pspacedelta}
\end{equation}

Substituting into (\ref{Omegamumu'}) and performing the $\pi'$-integration we
obtain:
\begin{eqnarray}
\int ds\,\langle
p|\Omega_{\mu\mu'}|p'\rangle&=&-16\pi^{2} G_{\mu\nu\mu'\nu'}\int\!\delta\pi\,
\delta\xi\,\delta\xi'\,ds\,
e^{i\,p\,\xi(s)}\, e^{
-ip'\xi'(s)}\dot\xi^\nu(s)\dot{\xi'}^{\nu'}(s)\, \delta^{jj'}\nonumber\\
&&a_{\xi'}(s)\left[\Omega_{\xi}T^{j}\, 
\Omega_{\xi}^{-1}\right]\left[\Omega_{\xi'}T^{j'}\, 
\Omega_{\xi'}^{-1}\right]\exp-i\int_0^{2\pi}dt\,\pi(t)\left[\xi(t)-\xi'(t)
\right]\nonumber\\ 
&&\qquad\qquad\qquad\qquad\int_{s_-}^{s_+}ds'\,ds''\,{{\beta_\epsilon(s'-s)
\beta_\epsilon(s''-s)}\over\pi_\alpha(s')\pi^\alpha(s'')}.
\end{eqnarray}
Integrating then over $\pi(\bar{s})$ for $\bar{s}> s_{+}$ and 
$\bar{s}< s_{-}$ yields up to factors:
\begin{eqnarray}
\int ds\,\langle
p|\Omega_{\mu\mu'}|p'\rangle =
&&\int\!\delta\xi\delta
\xi' ds\prod_{s_{-}<\bar{s}<
s_{+}}d^{4}\pi(\bar{s})\, G_{\mu\nu\mu'\nu'}
e^{i\left[p\,\xi(s)-p'\xi'(s)\right]}\nonumber\\
&&{{\left[\xi^{\nu}(s_{+})-\xi^{\nu}(s_{-})\right]
\left[\xi^{'\nu'}(s_{+})-
\xi^{'\nu'}(s_{-})\right]}\over{\left|\xi(s_{+})-\xi(s_{-})\right|
\left|\xi'(s_+)-\xi'(s_-)\right|}}\nonumber\\  
&&\left[\Omega_{\xi}T^{j}\, 
\Omega_{\xi}^{-1}\right]\left[\Omega_{\xi'}T^{j'}\,
\Omega_{\xi'}^{-1}\right] \prod_{\bar
s< s_{-},\, \bar s> s_{+}}\! \! \! \delta^4(\xi(\bar s)-\xi'(\bar s))
\nonumber\\
&&\int_{s_-}^{s_+}ds'\,ds''\,{{\beta_\epsilon(s'-s)\beta_\epsilon
(s''-s)}\over\pi_
\alpha(s')\pi^\alpha(s'')}\,
\, \delta^{jj'}\nonumber\\
&&\exp-i\int_{s_{-}}^{s_{+}}dt\,\pi(t)\left[\xi(t)-\xi'(t)\right]. 
\label{whatever}
\end{eqnarray}
We recall next that the $\Omega_\xi$-matrices in the above formula are 
actually what we called the ``hatted'' quantities and they depend on the
loop only for $\bar{s}\geq s_{+}$ and due to the $\prod\delta^4(\xi(\bar
s)-\xi'(\bar s))$ factor, we have $\xi=\xi'$ in this range, so that we can
replace the factor involving these $\Omega_\xi$-matrices with:
\begin{eqnarray}
\left[\Omega_{\xi}T^{j}\,\Omega_{\xi}^{-1}\right]\left[\Omega_{\xi'}T^{j'}\,
\Omega_{\xi'}^{-1}\right]\delta^{jj'}\, \, \, \rightarrow\, \, \, 
&&\left[\Omega_{\xi}T^{j}\,\Omega_{\xi}^{-1}\right]\left[\Omega_{\xi}T^{j'}\,
\Omega_{\xi}^{-1}\right]\delta^{jj'}\nonumber \\
&=&T^{j}T^{j},
\end{eqnarray}
where the last equality follows from the fact that $T^{j}T^{j}$ is a Casimir 
operator of the Lie algebra and therefore invariant under rotations in the 
Lie algebra. The result is a factor independent of $\Omega_\xi$ and of
$\xi$ being thus constant in the remaining integration, which is in fact 
the main reason why we changed right in the beginning to these so-called 
``hatted variables''.  Hence, since the exponentials in (\ref{whatever}) 
depend only on $\xi(\bar{s})$ and $\xi'(\bar{s})$ for $\bar{s}\in(s_{-},s_{+})$
and $\dot{\xi}^\mu(s)$, according to (\ref{xidotappr}), only on $\xi(s_+)$ and
$\xi(s_-)$, we can perform both the $\xi$- and the $\xi'$-integration  
for $0\leq \bar{s}\leq s_{-},\, \, s_{+}\leq \bar{s}\leq 2\pi$ by using the
relation:
\begin{equation}
\int\!\delta \xi\,\dot\xi^\nu(s)\,
   \dot\xi^{\nu'}(s) \propto {1\over 4}g^{\nu\nu'}\dot{\xi}^{2}(s).
\end{equation}

We now write: 
\begin{equation}
e^{i[p\,\xi(s)-p'\,\xi'(s)]}=\exp
i\int^{s_{+}}_{s_{-}}dt\,\Big{[}p\,\xi(t)-p'\xi'(t)\Big{]}\,
\beta_{\epsilon}(s-t),
\end{equation}
to give:
\begin{eqnarray}
\int\! ds\,\langle p|\Omega^{\mu\mu'}|p'\rangle\!\!&\!\!=\!\!&\!\!\,
\int\!\,
\prod_{s_-<\bar s< s_+}\,d^4\xi(\bar s)\,d^4\xi'(\bar
s)\,d^4\pi(\bar s)\,ds\,T^{j}T^{j'}\, \delta^{jj'}\nonumber\\
&&\int_{s_-}^{s_+}\! ds'\,ds''\,{{\beta_\epsilon(s'-s)\,
\beta_\epsilon(s''-s)}\over
{\pi_\alpha(s')\,\pi^\alpha(s'')}}g^{\mu\mu'}\nonumber\\
&&\exp
i\int_{s_-}^{s_+}\!dt\,[p\beta_\epsilon(s-t)+\pi(t)]\,\xi(t)\nonumber\\
&&\exp
-i\int_{s_-}^{s_+}\!dt\,[p'\beta_\epsilon(s-t)-\pi(t)]\,\xi'(t).
\end{eqnarray}
which can be simplified further to obtain:
\begin{equation}
\int\!ds\,\langle
p|\Omega^{\mu\mu'}|p'\rangle=g^{\mu\mu'}T^{j}T^{j'}\delta^{jj'}\, 
\delta^{4}(p-p')\frac{1}{p^{2}}.
\end{equation}
Our result for the $\widetilde{A}$-propagator in momentum space therefore is:
\begin{equation}
\widetilde{D}^{ii'}_{\mu\mu'}(p)T^{i}T^{i'}=g_{\mu\mu'}\delta^{ii'}\frac{1}
{p^{2}}T^{i}T^{i'},
\end{equation}
which is exactly what we wanted to prove.

\setcounter{equation}{0}
\section{Remarks}

Although the results we have obtained so far in attempting to extend the
discussion of monopole dynamics in Yang-Mills fields to the quantum theory
are quite limited, they have, we believe, given us some insight on several
points.

Firstly, the Wu-Yang criterion \cite{Wuyang} which has been applied in 
all previous work in the literature only to monopoles in the classical field 
theory, has now been shown to be extendable to the quantum field level to
define their dynamics and to generate Feynman diagrams.  In the nonabelian 
theory, the result cannot be checked, because the dynamics of monopoles is 
otherwise unknown.  However, the same calculation applies of course also to 
the abelian theory which is expected to be dual symmetric, so that the 
dynamics of monopoles there should be the same as that of ordinary charges.
Further, in the abelian theory, both the gauge boson self interaction term
and the Jacobian can be ignored so that our $g^0$ calculation given above is
exact.  Hence, the fact that we obtained the same ``perturbation series'' in
${\tilde g}$ above as the normal expansion in $g$ in ordinary electrodynamics
is a check not only on the Wu-Yang criterion but also of the loop space 
method employed.

Secondly, if the result recently obtained in the classical theory that 
Yang-Mills theory is dual symmetric \cite{Chanftsou1} is extendable to the 
quantum theory, one would expect that the dynamics of monopoles dealt with 
here, in spite of its original loop space formulation, should eventually be 
expressible in terms only of the {\it local} dual potential 
${\tilde A}_\mu(x)$.  It was seen that at least at the $g^0$ level 
we were working, this was indeed the case.  Whether it may persist at higher 
orders in $g$, and in such a way as to restore the dual symmetry, however, 
is at present unknown.

Thirdly, we have demonstrated that one can indeed do perturbation theory 
using the loop space techniques already developed.  The calculation is a 
little clumsy but perfectly tractable.  Though starting with loop variables 
which are invariant under the original $U$-transformation, it turns
out that in order to remove the intrinsic redundancy of loop variables, one
encounters in the Lagrange multiplier of the constraint a new field 
$L_{\mu\nu}[\xi|s]$ which depends on the dual (magnetic) ${\tilde U}$-gauge,
so that we had again to gauge-fix.  However, it is possible 
that by imposing the constraint in a different (global) manner 
\cite{Chanstsou}, one may have a chance of obtaining explicitly gauge 
invariant results.

For these reasons, in spite of the limited scope of the result obtained so far, 
we hope that it will serve as a basis for further explorations.

\section{Acknowledgements} 

JF acknowledges the support of the Mihran and Azniv Essefian Foundation 
(London), the Soudavar Foundation (Oxford) and the Calouste Gulbenkian 
Foundation (Lisbon), JP is grateful to Studienstiftung d.d. Volkes for 
financial support and TST thanks the Wingate Foundation for partial support
during most of this work.

\end{document}